\documentclass[sensors,article,accept,moreauthors,pdftex,10pt,a4paper]{./Definitions/mdpi} 
\firstpage{1} 
\articlenumber{1910}
\issuenum{8}
\pubvolume{19}
\pubyear{2019}
\copyrightyear{2019}
\history{Received: 12 February 2019; Accepted: 18 April 2019; Published: 22 April 2019}
\updates{yes} 

\usepackage{enumitem}
\usepackage{makecell}
\usepackage{multirow}
\usepackage{xhfill}
\usepackage{tabularx}

\Title{Data-Driven Interaction Review of an Ed-Tech~Application}

\Author{Alejandro Baldominos $^{1,2}$\orcidA{} and David Quintana $^{1,}$*\orcidB{}}

\AuthorNames{Alejandro Baldominos and David Quintana}

\address{%
$^{1}$ \quad Department of the Computer Science, Universidad Carlos III de Madrid, Leganés 28911, Spain; abaldomi@inf.uc3m.es\\
$^{2}$ \quad Smile and Learn Digital Creations, Madrid 28043, Spain}

\corres{Correspondence: dquintan@inf.uc3m.es; Tel.: +34-91-624-9109}

\abstract{\textls[-5]{Smile and Learn is an Ed-Tech company that runs a smart library with more that 100~applications, games and interactive stories, aimed at children aged two to 10 and their families. The~platform gathers thousands of data points from the interaction with the system to subsequently offer reports and recommendations. Given the complexity of navigating all the content, the library implements a recommender system. The purpose of this paper is to evaluate two aspects of such system focused on children: the influence of the order of recommendations on user exploratory behavior, and the impact of the choice of the recommendation algorithm on engagement. The~assessment, based on data collected between 15 October 2018 and 1 December 2018, required the analysis of the number of clicks performed on the recommendations depending on their ordering, and an A/B/C testing where two standard recommendation algorithms were compared with a random recommendation that served as baseline. The results suggest a direct connection between the order of the recommendation and the interest raised, and the superiority of recommendations based on popularity against other~alternatives.}}


\begin{document}

%
%
\section{Introduction and Background}
Smile and Learn's smart library is an application in the educational technology (Ed-Tech) space which is aimed at children. As of December 2018, the platform features a total of 107 games, which are grouped according to Gardner's theory of multiple intelligences \citep{Gardner83}. 

Contents, which include games, stories and videos, are designed to be used in different devices and rely on a common framework. The application registers thousands of data points as a result of user interaction. Based of these, the system can then generate personalized reports and recommendations relevant to users, parents and educators. 

\begin{figure}[H]
  \centering
  \includegraphics[width=\columnwidth]{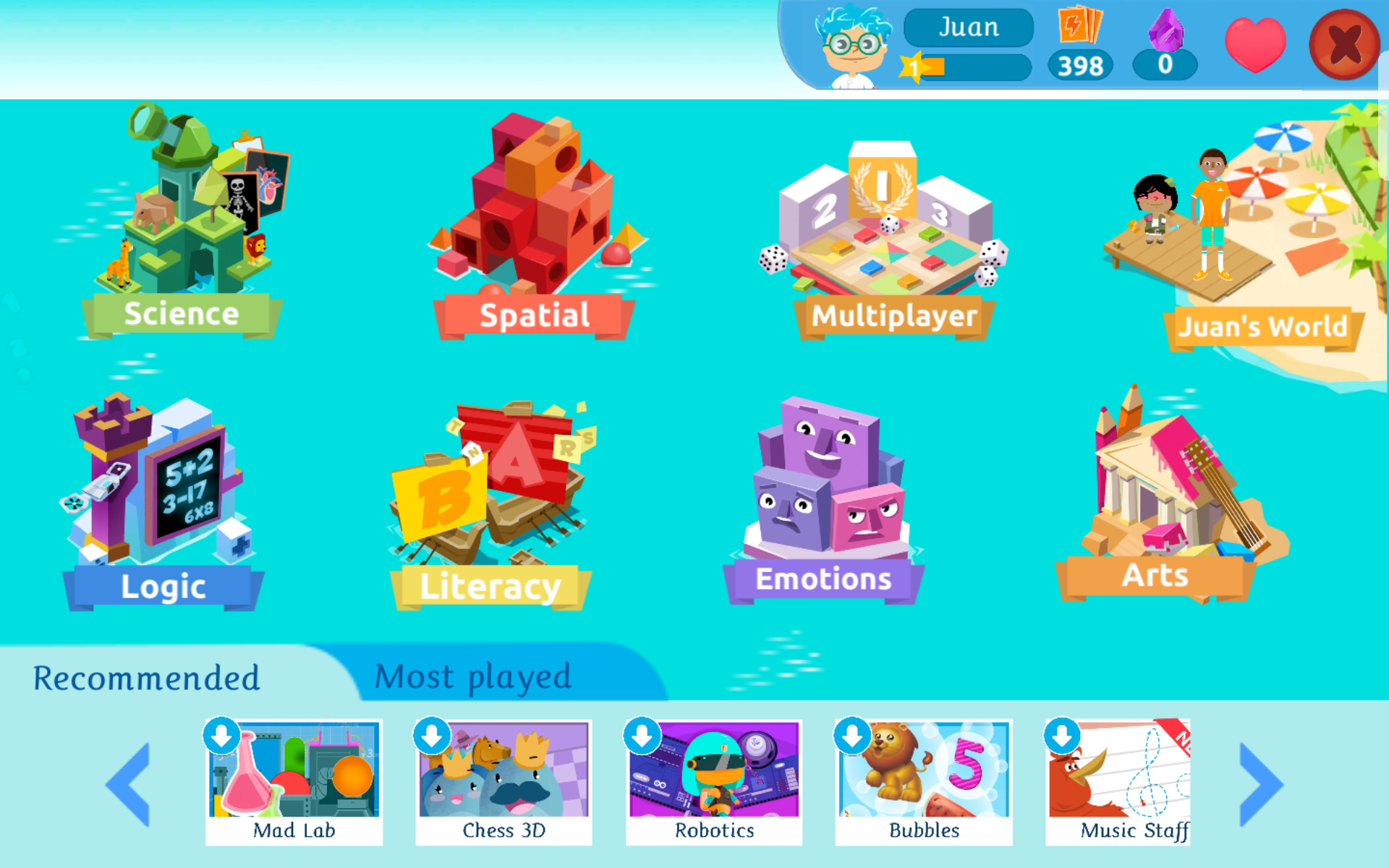}
  \caption{Screenshot of the main menu of Smile and Learn, showing the different worlds available in the application, each world corresponding to an intelligence.}
  \label{fig:mainmenu}
\end{figure}

The initial interaction with the application requires choosing among a large set of alternatives that are organized according to broad categories. These can be identified in Figure \ref{fig:mainmenu}.~The different games are grouped in so-called ``worlds'', with each world corresponding to an intelligence: science (naturalistic), spatial (visual-spatial), multiplayer (group-interpersonal), logic (logical-mathematical), literacy (verbal-linguistic), emotions (emotional-intrapersonal), and arts (artistic).~There is one additional world named after the user, which consists on a virtual village where the child interacts with characters to improve her wealth. Here, the dynamics require abilities from all intelligences.

\begin{figure}[H]
  \centering
  \includegraphics[width=\columnwidth]{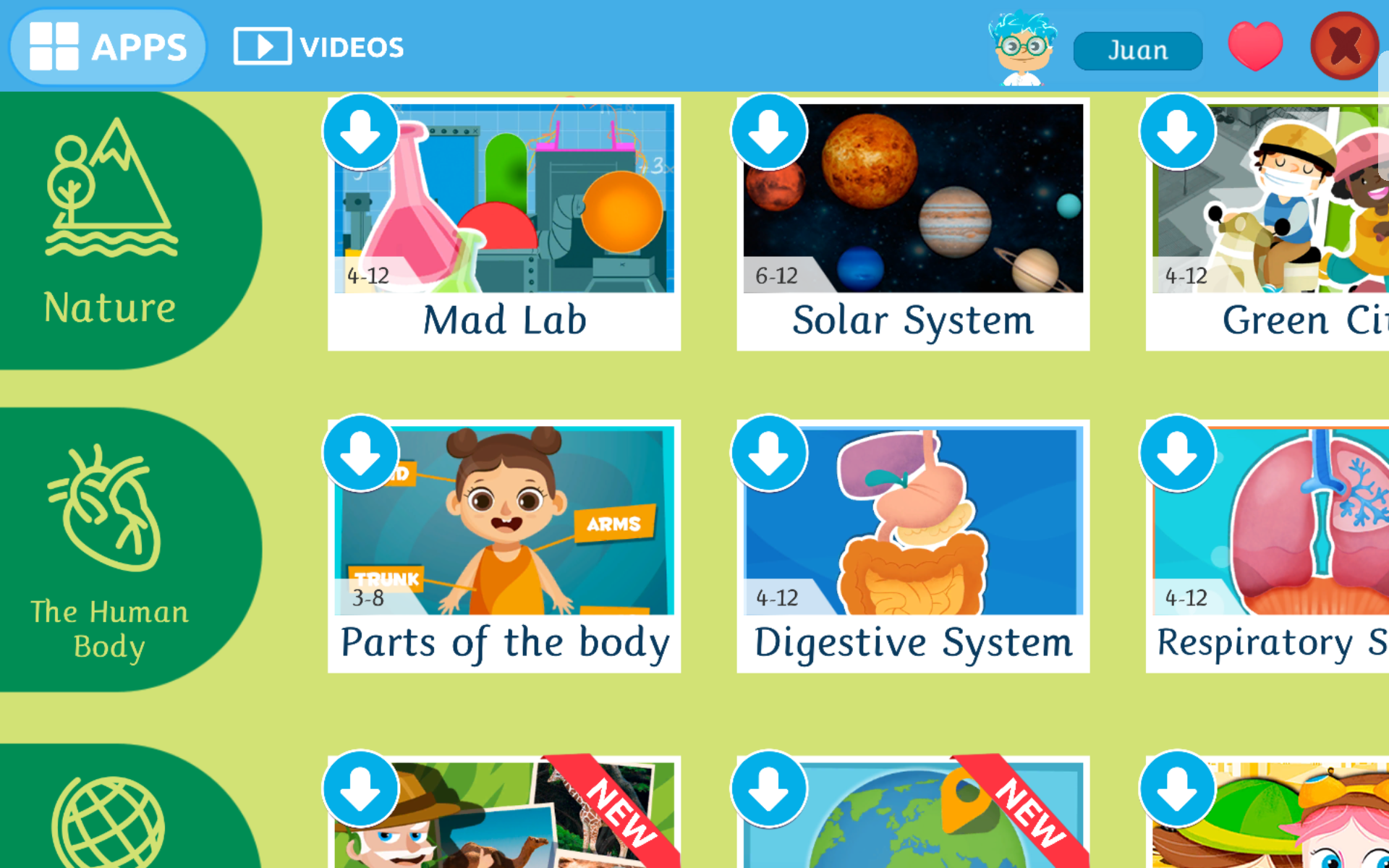}
  \caption{Screenshot of a world menu (in this case, the science world) in Smile and Learn, where apps belonging to the same world (or intelligence) are listed grouped by category.}
\label{fig:worldmenu}
\end{figure}

\textls[-5]{Once a world is chosen, apps are displayed grouped in categories, with one category per line. The~children can use the vertical scroll (swipe pattern) to navigate through categories, and the horizontal scroll to navigate between apps within a category. An example screenshot from the Science world is shown in Figure \ref{fig:worldmenu}.}

\unskip

Due to the increase in the number of games, which is still growing month after month, navigation through the application is becoming more cumbersome, meaning that it can take more time and effort to search for a certain game within the application. In order to alleviate this issue, we have introduced a navigation ribbon at the bottom of the main menu which allows a fast access to some chosen applications. As it can be seen at the bottom of Figure \ref{fig:mainmenu}, this navigation pattern consists of two tabs: ``recommended'' and ``most played''. The latter displays a list of the five top-played apps for the child using the app, turning into a useful mechanism for providing a fast access to those frequent apps. 

Regarding the former tab, it displays a list of up to seven apps recommended to children based on their usage behavior. The design motif behind this recommender system is not only to enable fast access to those apps which might be of interest for a child, but also to enhance exploration: letting children discover games which they might not reach if they had to find them by navigating through the~app.

The range of strategies developed to assign users to items is broad.~Some well established alternatives would be the recommendation of the most popular items~\citep{MHS07}; collaborative filtering~\citep{SFH07}; content-based approaches~\citep{BH04} or usage context-based similarity~\citep{NW13}. In addition to the basic possibilities, there is room for hybrid strategies that combine the output of several canonical strategies to generate their output.

Currently, two different strategies are implemented in Smile and Learn:

\begin{itemize}[leftmargin=*,labelsep=5.8mm]
    \item Popular: this approach recommends the most popular applications among other users. The~rationale behind this approach is that it is likely that some games are specially enjoyable for most children, because of their quality, design or playability, and therefore, they are safe to recommend, as there is a high chance that other children will find them engaging.
    \item Collaborative filtering: in this approach, the historical records of usage of all children are used in order to generate recommendations. The idea behind collaborative filtering is to find children similar to the one whose recommendations are being generated, so that these recommendations consist of games that have been played by these similar children, but not the child being targeted.
\end{itemize}

Both these strategies are based on implicit feedback, since explicit feedback given by children can turn out to be unreliable or difficult to interpret. Out of the different implicit parameters that could be taken into account \cite{Rolando15}, we have used the number of games played by child and app, as well as the duration of such games.

The specifics of the implementation of these algorithms are discussed in relation to a prior version of the system by \citet{RMB18}.

In the navigation bar displayed at the bottom of the main menu, recommendations from different recommenders could be combined. Also, it is worth noting that while there is a maximum of seven recommended games, in some screen sizes or formats it can occur that less than seven are displayed and then the list becomes scrollable. Based on our experience, most screens are able to show five recommendations, with the last two being hidden and accessible only upon scroll (through a swipe pattern or by pressing the side arrows, an example can be seen in Figure \ref{fig:mainmenu}).

In this paper, we want to design a sound experiment in order to determine which recommender strategy is working best, i.e., which one is translating into a higher engagement of children with the recommended games. We are also interested in getting insights about the interaction patterns taken by children when dealing with the recommendations. With this information, we could improve the system and enhance user experience.

This work contributes valuable new evidence on the reaction of children to two well-established methods of generating recommendations, and assessing the importance of the order of presentation. Information like this, specifically focused on children in the educational space, is both scarce and very hard to obtain. Given that the related literature on children is very limited, we  consider that the results might be very relevant to developers of K-12 Ed-Tech applications.

The rest of the document is structured as follows: in Section \ref{sec:relwork} we present related works on recommender systems and interaction patterns within the context of an Ed-Tech application. That~will be followed by Section \ref{sec:methods}, where we describe the experimental methodology. The results are then reported in Section \ref{sec:results}. Finally, conclusive remarks are presented in Section \ref{sec:concl} along with some suggested future lines of work.

\section{Related Work}
\label{sec:relwork}

Applications of technology to educational processes (a field known under the term ``Ed-Tech'') are not particularly new. However, it has only been in the last decade that the availability and ubiquity of technology devices (such as smart-phones or tablets) has reached a point in which these applications can be deployed in large-scale settings.

Additionally, the implementation of data science or machine learning techniques are allowing to extend Ed-Tech far beyond the classical definition of using technology to deliver contents to an audience (e.g., slides, interactive videos, etc). \citet{IJIMAI-1452} discuss how novel development allow tracking the progress of individual users, detecting strengths and weaknesses, and providing a customized experience to enhance the learning process, among other possibilities. 

Many examples of such novel applications can be found in the literature in recent years. For~example, Charleer et al.~\cite{Charleer17} presented a learning analytics dashboard aimed at improving the communication in advising sessions, helping to increase students' motivation. Lange et al.~\cite{IJIMAI-2126} also made related contributions in the space of virtual training centers and \citet{Kaser17} have proposed an approach to student modelling to represent and predict students' knowledge and skills. Some authors have been working on the prediction of academic performance 
\citep{IJIMAI-2114} 
and others, like \citet{UM17} introduced a system designed to scaffold learning in specific domains such as programming. 

An important subset of Ed-Tech applications are those aiming at providing a customized experience by suggesting users a certain topic or learning process. This problem can generally be regarded as a recommendation problem. Recommender systems, the instruments used to tackle the problem, have a three-decade long history and have been the subject of a sizable amount of research. These efforts on recommender systems in general have been surveyed by \citet{BOH13} and, more recently, by \citet{LWM15}. If we focus specifically in education, the literature reviews authored by \citet{DVS15} and \citet{EFR15} are specially insightful and comprehensive.

The design and implementation of recommender systems in the education space poses a number of difficulties, as discussed by \citet{TN17}, and so does their evaluation~\citep{LM15}. Among these challenges, we can mention the selection of the right algorithm (or set of algorithms) and the design of the interface. Even though the first aspect is not settled yet, there have been valuable efforts to contribute evidence in this regard like a recent work by \citet{KKL16}. 

These authors cite among the best-established and computationally inexpensive tag and resource recommendation strategies for technology-enhanced learning: base level learning equation with associative component \citep{KKS15}; collaborative filtering \citep{SFH07}; content-based \citep{BH04}; most popular \citep{MHS07}; SUSTAIN~\citep{LMG04} or usage context-based similarity \citep{NW13}.

\textls[-5]{In this regard, studies like \cite{PAC17} offer systematic reviews of different algorithmic alternatives, while~\cite{DMK08,MVD11} provide specific lists of advantages and disadvantages of the main recommendation techniques for technology-enhanced learning.~These are briefly summarized in Table \ref{tab:algorithms}, which adapts a very similar one included in the last two mentioned studies, adding popularity-based recommendations to the set of techniques.}

Besides the alternatives displayed in the table, an additional category of recommender systems are hybrid approaches. In this case, two or more of these techniques are fused in order to alleviate some of the disadvantages of a particular recommender. This possibility was discussed in detail in a survey on Next Generation of Recommender Systems by \citet{AA06} and, later by \citet{AB2015}.

\begin{table}[H]
  \centering
  \caption{Characteristics of main recommendation techniques suitable for technology-enhanced learning. CF: collaborative filtering. Adapted from \citet{DMK08} and \citet{MVD11}.}
    \begin{tabular}{>{\raggedright\arraybackslash}p{2.5cm}>{\raggedright\arraybackslash}p{3.25cm}>{\raggedright\arraybackslash}p{3.25cm}>{\raggedright\arraybackslash}p{3.25cm}}
    \toprule
    \textbf{Technique} & \textbf{Description} & \textbf{Advantages} & \textbf{Disadvantages}\\
    \midrule
    Popularity & Recommends items that are popular among all users.\linebreak & No new user problem.\linebreak No content analysis. \linebreak Domain-independent. & Only popular taste.\linebreak Insensitive to changes of preferences.\\
   \midrule
    User-based CF & Users who rated the same item similarly probably have the same taste. Based on this assumption, this technique recommends the unseen items already rated by similar users.\newline& No content analysis.\newline Domain-independent.\newline Quality improves.\newline Bottom-up approach.\newline Serendipity. & New user problem.\newline New item problem.\newline Popular taste.\newline Scalability.\newline Sparsity.\newline Cold start problem.\\
    \midrule
    Item-based CF & Focus on items, assuming that the items rated similarly are probably similar. It recommends items with the highest correlation (based on ratings for the items).\newline& No content analysis.\newline Domain-independent.\newline Quality improves.\newline Bottom-up approach.\newline Serendipity. & New item problem.\newline Popular taste.\newline Sparsity.\newline Cold start problem.\\
   \midrule 
    Stereotypes or demographics CF & Users with similar attributes are matched, then it recommends items that are preferred by similar users (based on user data instead of ratings).\newline & No cold start problem.\newline Domain-independent.\newline Serendipity. & Obtaining information.\newline Insufficient information.\newline Only popular taste.\newline Obtaining metadata information.\newline Maintenance ontology.\\
    \midrule
    Case-based reasoning & Assumes that if a user likes a certain item, she or he will probably also like similar items. Recommends new but similar items.\newline& No content analysis.\newline Domain-independent.\newline Quality improves & New user problem.\newline Overspecialisation.\newline Sparsity.\newline Cold start problem.\\
    \midrule
    Attribute-based techniques & Recommends items based on the matching of their attributes to the user profile. Attributes could be weighted for their importance to the user. & No cold start problem.\newline No new user/new item problem.\newline Sensitive to changes of preferences.\newline Can include non-item-related features.\newline Can map from user needs to items. & Does not learn.\newline Only works with categories.\newline Ontology modelling and maintenance is required.\newline Overspecialisation.\\
    \bottomrule
    \end{tabular}%
  \label{tab:algorithms}%
\end{table}%

\scalebox{.945}[1.0]{ Among the works that could illustrate this possibility, we could mention one by Rodriguez~et~al.~\cite{Rodriguez15}}, \textls[-5]{who built a hybrid recommender combining content-based, collaborative filtering and knowledge-based approaches to develop a recommender system that suggests learning objects extracted from a repository in an educational setting. Another example can be found in the work by \citet{Salehi12}, which combines content-based and collaborative filtering recommenders, in this case to suggest appropriate learning materials.~\citet{BO18} introduce a recommender system for web-based education that adapts its recommendations of learning objects to the learning styles of users. In this case, the hybrid recommender system is based on collaborative filtering and association rule mining. Finally, \citet{NFH18} recently introduced the ULEARN system, that also selects and sequences learning objects that match the learning styles of the users. To~this end, they rely on a hybrid recommendation approach that includes collaborative filtering and content~filtering.}

The second aspect is specially complicated when it involves children, as their cognitive development makes their interests and capabilities change dramatically over a period of relatively few years \citep{sears2009}, and the fact that intuitions by adult designers might not be correct~\citep{Bjorklund17}. Even though there have been advances in this aspect, like the contribution of \citet{Wu2014} on interface design for children, there is still a significant amount of work to be done. 

If we focus on recommendation for children, even though we could mention some relevant works like that by \citet{PN14} the volume is still very scarce. To illustrate this, it is worth mentioning that the first specialized workshop, {KidRec}, took place in 2017~\citep{PFG18}. As \citet{Deldjoo2017enhancing} explain, recommender systems have been traditionally focused on adults and, when it comes to children, the field is still in its infancy.

\section{Materials and Methods}
\label{sec:methods}
In this study we intend to compare the performance of two popular recommender strategies to suggest potential games of interest to children. As it was described earlier, these are providing recommendations based on apps commonly played by most children, and collaborative filtering, whose recommendations are based on games played by similar users, i.e., those who have a similar record of played games.

In order to get a better understanding of which strategy was working best, we tested the performance of both recommender strategies and, additionally, we compared them against a baseline recommender which made random recommendations. Given that the aim of the work in this regard was descriptive, not prescriptive, an A/B/C testing, a well-established evaluation approach in this context, was performed. This methodology was recently used by \citet{KRG18} to evaluate their recommender algorithm in Taobao display advertising platform in production, and so was used by \citet{HA18} to evaluate theirs on a large scale online test on an online video portal.

During the whole period in which the experiment was running, each child was assigned one group (either A, B or C) randomly following a uniform distribution. When running the recommender, children received recommendations coming from a different strategy based on their group:

\begin{enumerate}[label=\Alph*.]
    \item Popular strategy was used.
    \item The collaborative filtering strategy was used.
    \item The random strategy was used.
\end{enumerate}

Instead of generating all the recommendations (a maximum of seven) from the resulting strategy, only the first three recommendations were computed using the recommender corresponding to the child's group, and the remaining recommendations were chosen randomly.

In all cases, some filters were applied in order to opt out of some recommendations which might not be suitable:

\begin{itemize}[leftmargin=*,labelsep=5.8mm]
    \item Some games were blacklisted, meaning that they might not have enough quality as to be recommended (e.g., they were in a beta stage).
    \item Games were filtered out if they were not available in the version of the app owned by the user (e.g., a game was introduced in version 4 and the child is using version 3).
    \item Games were filtered out if they were not designed for the range of age of the target child (e.g., a game is aimed at children 4--6 years old, but the child is 3 years old).
\end{itemize}

It is worth mentioning that these filters were applied in all cases, even in the random recommendation strategy. 

\subsection{Algorithms}
As stated earlier, three different recommendation strategies were followed. One of these strategies was entirely random, and was used as a baseline. Its implementation was trivial, as apps to be recommended were randomly sampled from among the whole set.

\subsubsection{Popular Recommender}

The second strategy was the popular one. For the popular recommender, the algorithm computed a metric of normalized interest, $\overline{I}(a)$, for each app $a$, which was immediately derived from the number of games ($\#Games(a)$) and the amount of time in days that the app has been available in the library ($Age(a)$), as shown in Equation (\ref{eq:pop}):
\begin{equation}
    \label{eq:pop}
    \overline{I}(a) = \frac{\#Games(a)}{Age(a)}.
\end{equation}

Once the normalized interest was computed for each app, then all apps were ranked according to this value and the top-$k$ apps were returned by the recommender.

Two design decisions about the popular recommendation system are worth mentioning. First, only the amount of games played for an app was taken into account, while the duration of these games was ignored. This was done on purpose, as each app has its own particularities and some might naturally lead to longer games than others, and we did not want to set these apps higher in the ranking. However, it must be noted that games with a duration of less than five seconds have been ignored. Second, the interest was normalized by the amount of time that the app has been published in the library. While another alternative could have been chosen (for example, considering only the data from the last $d$ days), we have intentionally performed this normalization so that new apps had a higher chance to be chosen by the recommender.

\subsubsection{Collaborative Filtering Recommender}

The third recommendation strategy was based on collaborative filtering. This strategy relies on implicit rather than explicit feedback. This decision was motivated by the fact that small children have their own interacting patterns, and might be unable to properly give credit or score to a game after playing it \citep{Deldjoo2017enhancing}. For this reason, a system based on explicit feedback could be unreliable. Instead, we are basing recommendations on implicit patterns of interaction, and in particular in how much and how long a child played with a game.

The idea behind collaborative filtering was to recommend apps to a child which can be of high interest based on the preferences of other children which are deemed similar. Therefore, our implementation of collaborative filtering was done in two stages. In the first stage, the neighborhood was formed by choosing children similar to the one to whom recommendations are being provided. In~the second phase, a rating was predicted for each candidate app based on the neighborhood and the top-$k$ apps were selected for the recommendation.

In the neighborhood formation stage, for each target child to whom recommendations were provided, we needed to select similar children on which the recommendations relied.~First of all, only children with at least one game of the same application in common were considered for the neighborhood (or equivalently, children whose set of played apps was disjoint with that of the target child are systematically excluded for the neighborhood). Each child's profile included some demographic data including the age and the gender. In this case, the age was used to form the neighborhood; however, the gender was intentionally omitted to prevent unintended biases.

To establish the neighborhood, a similarity metric between two children $c_1$ and $c_2$ was defined as shown in Equation (\ref{eq:sim}):
\vspace{12pt}
\begin{equation}
    \label{eq:sim}
    Sim(c_1,c_2) = 0.4 \cdot Sim_{age}(c_1, c_2) + 0.6 \cdot \frac{\left|Apps(c_1) \cap Apps(c_2)\right|}{\left|Apps(c_1) \cup Apps(c_2)\right|}.
\end{equation}

As it can be seen, this similarity measure relied on a weighted average of two different criteria: the similarity in the age and the similarity of the gaming history of the two children, and returned a value in the range $[0,1]$. In the case of the former criterion, it was computed following Equation (\ref{eq:age}):
\begin{equation}
    \label{eq:age}
    Sim_{age}(c_1, c_2) = 
  \begin{cases} 
   1    & \text{if } Age(c_1) = Age(c_2), \\
   0.5  & \text{if } |Age(c_1) - Age(c_2)| = 1, \\
   0    & \text{otherwise}.
  \end{cases}
\end{equation}

In the previous equation, $Age(c)$ refers to the age of child $c$ in years. The previous function has been designed to fit an educational reality: a difference of more than one year in such an early stage of the educational process (two to 10 years old) was very large, and two children with such an age gap were likely be very different.

The latter part of Equation (\ref{eq:sim}) refers to the similarity in terms of the used apps, which was the main criterion for the implicit feedback collaborative filtering.~This was computed as the intersection-over-union of the sets of apps played intensively by the two children. By ``intensively'' we refer to apps that have a cumulative of at least 10 games and 60 s of game. This filtering has been done to prevent considering apps which might been rarely played by a children while examining or exploring Smile and Learn library. 

Neighborhood formation was carried out by choosing the top-100 children with higher similarity. A minimum similarity of 0.5 was enforced in order for a child to be included in the neighborhood.

Finally, for each candidate app $a$ to be recommended we computed the average of the ratings $r(n,a)$ for each children $n$ in the neighborhood $N$. Since we wanted to deal with implicit feedback, this rating corresponds to the number of games of children $c$ in app $a$. The average of ratings was weighted by the similarity of the target children and each neighbor. In summary, the interest of children $c$ in app $a$, $I(c, a)$ was computed as shown in Equation (\ref{eq:int}):
\begin{equation}
    \label{eq:int}
    I(c,a) = \frac{\sum_{n \in N} \left(Sim(c,n) \cdot r(n,a) \right)}{|N|}.
\end{equation}

Finally, apps were sorted by their interest and the top-$k$ apps were provided as recommendations.

\subsection{Dataset}
\label{ssec:dataset}

During the period of the experiment, from 15 October 2018 to 1 December 2018, 
we recorded the following information, which we used to later evaluate the system and report the results:

\begin{itemize}[leftmargin=*,labelsep=5.8mm]
    \item The recommender (popular, collaborative filtering, or random) assigned to each child.
    \item The recommendations generated, and also which recommender provided each of these recommendations.
    \item The date and time at which each recommendation is generated.
    \item The games usage per child, including the times at which they play and the duration for each~game.
\end{itemize}

It is worth noting that regarding the second aspect, recommendations might not always be generated using the desired recommender. For example, a child might be assigned the collaborative filtering strategy, but this recommender might not have enough information about usage as to generate useful recommendations, or that all of these recommendations are filtered out. In that case, random recommendations are provided instead.

With this data, we can explore the patterns of engagement of children with the different games depending on whether those games were recommended or not.

\subsection{Performance Metrics}
\label{ssec:metrics}

The impact of the position in the ribbon of recommendations was be analyzed using click-through frequencies. That way, we will determined whether the first five visible items got more clicks than the last two, and whether the ordering within the visible and invisible ones mattered.

\textls[-5]{The core assessment of the recommendation algorithms will be made according to two engagement metrics. One based on the number of games and another one on game time.}

The first one, the average number of games per user (ANG), is formally defined in Equation (\ref{eq:ang}).
\begin{equation}
\label{eq:ang}
	ANG= \frac{\sum_{i=1}^{NumUsers_{R}}Games_{Ri}}{NumUsers_{R}},
\end{equation}
where $Games_{Ri}$ is the number of games played by user $i$ on apps recommended by algorithm $R$ (either {collaborative filtering, popular} or {random}, and $NumUsers_{R}$ is the total number of users who were recommended the apps by algorithm $R$ and acted on it.

The second aspect of engagement to be measured was the average game time ($AGT$) by users who acted on the recommendations. The expression used to compute the indicator is described in Equation~(\ref{eq:agt}).
\begin{equation}
\label{eq:agt}
	AGT = \frac{\sum_{i=1}^{NumUsers_{R}}GameTime_{Ri}}{NumUsers_{R}},
\end{equation}
where $GameTime_{Ri}$ is the total time spent by user $i$ playing apps recommended by algorithm $R$ (either {collaborative filtering, popular} or {random}, and $NumUsers_{R}$ is the total number of users who were recommended the apps by algorithm $R$ and used them. 

We should note that, for the computation of these metrics, we define ``game'' as the event where the user interacts with the application for 10 s or more. Given the presence of some outliers, we also filtered out games of more than 3000 s and the instances where a game was played more than 60 times by the same user. These accounted for less than 0.5\% of the sample.

In order to provide a more complete picture, in addition to these engagement metrics, we reported four standard performance indicators: accuracy, precision, recall and F1 score.

To compute these metrics we have proceeded as follows: we have divided the pilots period, comprising a total of 45 days, into two different time spans.~The first one comprised one month (from 15 October 2018 to 14 November 2018) and the second comprised 15 days (from 15 November 2018 to 1 December 2018).~The former was used as the training set while the latter was used for validation~purposes. 

We have considered the apps suggested by the recommender during the training phase to each child and have then matched this information with the actual apps played by children during the testing period. With this information, we have been able to build a confusion matrix as follows: apps recommended to a child during training and then played during testing constituted a true positive ($TP$), apps recommended but not played counted as false positives ($FP$), apps played but not previously recommended were considered false negatives ($FN$) and apps not recommended and not played constituted the true negatives ($TN$). All confusion matrices for each child were summed up for each recommendation strategy, and then accuracy, precision, recall and F1 score were computed following their standard definitions, which are shown in Equations (\ref{eq:acc})--(\ref{eq:f1}).
\begin{equation}
    \label{eq:acc}
    Accuracy = \frac{TP + TN}{TP + TN + FP + FN},
\end{equation}
\begin{equation}
    \label{eq:prec}
    Precision = \frac{TP}{TP + FP},
\end{equation}
\begin{equation}
    \label{eq:rec}
    Recall = \frac{TP}{TP + FN},
\end{equation}
\begin{equation}
    \label{eq:f1}
    F1\:Score= 2\cdot\frac{Precision \cdot Recall}{Precision + Recall}.
\end{equation}

\section{Results and Discussion}
\label{sec:results}

As it was mentioned before, the first part of the study has to do with the analysis of the impact of the position in the recommendation ribbon on the click-through. During the pilots period we have measured a total of 30,516 clicks received by 89 different apps, which have been involved in a total of 472,498 games. The experimental results on this aspect are reported in Table \ref{clickthrough}. There, we can see the total number of clicks over the relevant period by position for the apps that were recommended in all the possible slots.

\begin{table}[H]
  \centering
  \caption{Accumulated click-through by position in the recommendation ribbon and mean for the visible and invisible positions for the period from 15 October 2018 to 1 December 2018.} 
    \begin{tabular}{ccccccccc}
    \toprule
    \multirow{2}[2]{*}{\textbf{Position}} & \multicolumn{5}{c}{\textbf{Visible}} & & \multicolumn{2}{c}{\textbf{Hidden}} \\\cmidrule{2-9}
          & \textbf{1} & \textbf{2} & \textbf{3} & \textbf{4} & \textbf{5} & & \textbf{6} & \textbf{7} \\
    \midrule
    Clicks & 6329 & 5834 & 4516 & 2639 & 3960 &  & 3150 & 2937 \\
    Mean & -&-& 4830 &- & -&& 3044 &- \\
    Rank & 1 & 2 & 3 & 7 & 4 &  & 5 & 6 \\
    \bottomrule
    \end{tabular}%
  \label{clickthrough}%
\end{table}%

It is apparent that the first three positions grabbed much more interest than the rest. The difference between the first two and the other five were specially sizable. While it was true that the fourth received fewer clicks than the ones that follow, other than that, the results were consistent with the existence of a direct relationship between the order and the number of times that users acted on the recommendations.

If we consider visibility, the ribbon only showed five recommendations at a time. We expected that to be a relevant factor, as getting to the last two requires a supplementary effort from the user. Interestingly, even though the average number of clicks on the visible slots was 4830, higher than the 3044 average clicks on the hidden ones, the role of friction as an element that drags click-through down could be questioned. If we consider the effect of position, where apps on the left hand side gather more interest that those on the right, and we infer the trend line, the recommendations on the sixth and seventh positions get more clicks than expected.\enlargethispage{0.5cm}

Regarding engagement, a total of 12,229 games were completed in apps that were being recommended at the moment by a total of 1387 children, summing up a total of 598 gaming hours. From the study of engagement, we find that the recommender algorithm based on popularity was superior across metrics. As we can see in Table \ref{globalresults}, it outperformed the other two both in terms of number of games and accumulated use time by user. Unexpectedly, the proposed collaborative filtering algorithm resulted in a slightly lower mean number games vs. the random alternative. The sign of this difference, however, was the opposite for game time.

\begin{table}[H]
  \centering
  \caption{Engagement metrics for game apps by recommender for the period from 15 October 2018 to 1 December 2018.} 
    \begin{tabular}{llccc}
    \toprule
    \multicolumn{2}{c}{\textbf{Engagement Metric}} & \multicolumn{1}{c}{\textbf{Mean}} & \multicolumn{1}{c}{\textbf{Median}} & \multicolumn{1}{c}{\textbf{Variance}} \\
    \midrule
    Games &       &       &       &  \\\midrule
          & $NG_{Random}$ & 4.79 & 3 & 34.22\\ 
          & $NG_{Popular}$ & 6.72 & 4 & 53.89\\
          & $NG_{CF}$ & 4.22 & 2 & 24.75\\
          \midrule
    Time  &       &       &       &  \\\midrule
          & $GT_{Random}$ & 740.98 & 405.26 & 1,084,308.47\\
          & $GT_{Popular}$ & 1202.76 & 639.79 & 3,286,849.96\\
          & $GT_{CF}$ & 776.01 & 295.37 & 5,020,886.62\\
    \bottomrule
    \end{tabular}%
  \label{globalresults}
\end{table}%

The statistical significance of the differences reported in Table \ref{globalresults} was assessed according to the protocol that follows. First, we started testing the normality of the distribution of the engagement metrics with Kolmogorov--Smirnov test with Lilliefors correction. In case normality was rejected, we~applied Wilcoxon's test. Otherwise, we tested for the presence homoskedasticity using Levene test and, based on the result, we relied either on Welch test, or the traditional t-test. The results were analogous for the metrics. The superiority of the Popular algorithm over the other two was significant at 1\%. Regarding the comparison of the baseline vs the implementation of collaborative filtering, the null hypothesis of equality could not be rejected at the 5\% conventional level for neither the number of games nor the total game time.

In regards to the standard performance metrics, we report the main ones (accuracy, precision, recall and F1 Score) in Table \ref{globalperresults}. As we can see, recommendations based on popularity offered the largest precision and F1 score, while collaborative filtering and random recommendations provided the highest values in terms of accuracy and recall, respectively.

\begin{table}[H]
  \centering
  \caption{\textls[-15]{Performance metrics for game apps by recommender algorithm. Train: 15 October 2018 to 14 November 2018}. Test: 15 November 2018 to 1 December 2018.}
    \begin{tabular}{lccccc}
    \toprule
    \textbf{Algorithm} & \textbf{Accuracy (\%)} & \textbf{Precision (\%)} & \textbf{Recall (\%)} & \textbf{F1 Score (\%)}\\
    \midrule
    Col. Filtering & 94.92 & 7.40 & 1.28 & 2.18 \\
    Popular & 94.73 & 12.93 & 3.30 & 5.25 \\
    Random & 93.51 & 6.64 & 3.57 & 4.64 \\
    \bottomrule
    \end{tabular}%
  \label{globalperresults}
\end{table}%

These results are influenced by the nature of the algorithms and the variety or recommendations that they generate. At the ends of the spectrum we would have random recommendation and recommendations based on popularity. The former approach is less constrained than the other two and, as a result, it foments exploration to a higher extent. Among the remainder, even though both are more stable, collaborative filtering provides a richer range of alternatives, as unguided exploration by peers is likely to result in new app recommendations. 

A second aspect to be considered regarding the performance indicators is that, unlike the engagement ones, they only provide information on whether the user felt compelled to try new apps, not whether the user liked them. We are more interested in discovering this latter extent, which we find more important in the scope of this study. For this reason, we find that the average number of games and average time spent playing the recommended apps are the key indicators to consider, whereas standard performance metrics can be useful supplemental information to gain a better understanding of the situation.

To summarize the results, the two elements that constituted the subject of study of the recommender system were the influence of the order of recommendation on user exploratory behavior and the impact of the choice among two well-established recommendation algorithms, pre-selected by the company, on engagement. In order to evaluate such aspects, we acquired data of real interaction with the application between 15 October 2018 and 1 December 2018, running an A/B/C test with three different implementations of a recommender system (including a random baseline) and measuring clicks made on recommendations.

We found a direct association between the order in the recommendation ribbon and the number of clicks. Apps on the left gather more interest that those on the right hand side. The friction introduced by the fact that reaching the last recommendations requires either swiping or pressing on the arrows on the side does not seem to have a negative impact. 

The A/B/C testing analysis used to compare the two recommendation approaches, one that recommends apps based on popularity and an implementation of collaborative filtering, vs the random recommender used as baseline offered two main results in regards to engagement: The first one is that the popular algorithm beats the other two both in terms of number of games and the total time spent playing the recommended contents. The second is that the current implementation of collaborative filtering does not seem to add value, as it offers the same performance as the baseline.~A~likely explanation to this outcome is that there are some games that due to their nature are very likeable, and for this reason they end up being ``the popular ones''. These games will end up being suggested by the recommender based on popularity, and if children start to play them, in most cases they will quickly become engaged. This process could be adding some bias in the computed engagement metrics in favor of the popular recommendation strategy. If we consider performance indicators, the approach based on popularity seems to be the best alternative in terms of precision and $F$1 score, but collaborative filtering offers the highest accuracy.

\section{Summary and Conclusions}
\label{sec:concl}

In this paper we have described and evaluated the behavior of a recommender system in the scope of an Ed-Tech application aimed at children aged 2--10 and their families. The smart library gathers thousands of data points based on user interaction and uses that information to generate tailored reports and recommendations. The latter aspect is managed by a recommender system designed for the purpose of easing navigation through the library and enhancing exploration. We evaluated two aspects of the system: the influence of the order of recommendations on user exploratory behavior, and the impact of the choice of the recommendation algorithm on engagement.

The analysis led to the following conclusions: first, the order in the recommendation ribbon matter. Apps in the left positions got more clicks, and the effort to access hidden recommendations did not seem to have a negative impact. The second one is that the popular algorithm resulted in more engagement than the other two. Finally, the current implementation of collaborative filtering does not seem to add value. Future works would include thorough studies on alternatives to the current implementation of collaborative filtering or the optimization of its parameters; studying new possibilities based on the interface to drive the attention of the user to the recommended applications; or the implementation of recommendation strategies based on competences aimed at fostering the development of the user according to predefined preferences established by parents and educators.

\vspace{6pt} 

\authorcontributions{D.Q. surveyed the state of the art; A.B. and D.Q. conceived and designed the experiments; A.B. performed the experiments, A.B. and D.Q. analyzed the data; A.B. and D.Q. wrote the paper.}

\funding{This research has received funding from the European Union's Horizon 2020 Research and Innovation Programme under grant agreement No. 756826.}


\conflictofinterests{The Authors declare no conflict of interest.} 



\reftitle{References}

\end{document}